\documentclass[letterpaper]{article}
\usepackage{aaai24}  
\usepackage{subcaption}
\usepackage{times}  
\usepackage{helvet} 
\usepackage{courier}  
\usepackage[hyphens]{url}  
\usepackage{graphicx} 
\usepackage{booktabs}
\usepackage{natbib}  
\usepackage{caption} 
\usepackage{multirow}
\usepackage{subcaption}
\usepackage{duckuments}
\usepackage{enumitem}

\usepackage{xcolor}
\newcommand{\answerYes}[1]{\textcolor{blue}{#1}} 
\newcommand{\answerNo}[1]{\textcolor{teal}{#1}} 
\newcommand{\answerNA}[1]{\textcolor{gray}{#1}} 

\newcommand{\change}[1]{\textcolor{black}{#1}} 

\begin{document}

\title{On the Effectiveness of Fact Checking Information from Politically Congruent and Incongruent Large Language Models}

\author {
    Jiangen He\textsuperscript{\rm 1*}, Benjamin D. Horne\textsuperscript{\rm 1*}, and Dorit Nevo\textsuperscript{\rm 2}
}
\affiliations {
    \textsuperscript{\rm 1} School of Information Sciences, University of Tennessee Knoxville, Knoxville, TN, USA\\
    \textsuperscript{\rm 2} Lally School of Management, Rensselaer Polytechnic Institute, Troy, NY, USA\\
    jiangen@utk.edu, bhorne6@utk.edu, nevod@rpi.edu\\
    \small{\textit{* indicates shared first-authorship and equal contributions}}\\
}

\maketitle

\begin{abstract}
\change{Social media companies have shifted away from human fact-checkers and instead have embedded conversational Large Language Models (LLM) on their platforms. LLM chatbots differ from human fact-checkers in many ways that may shape user responses to corrections. Of particular interest in this study is that LLM chatbots can be ideologically configured via the content emphasized in their responses, the sources cited, and the configured persona.} Using data from two within-subjects experiments (n=705), this paper investigates the effectiveness of fact checking information from ideologically configured LLM chatbots. We find that LLM fact-checkers significantly shift trust in true and false political news headlines, even when the chatbot is politically incongruent with the user. \change{The perceived political congruency between the participant and the bot matters only when headlines are politically distant. That is, trust in correctly labeled true headlines increases less when politically distant chatbots check distant headlines and increases more when moderate chatbots check distant headlines. The perceived political congruency of LLM chatbots did not impact their effectiveness at decreasing trust in false headlines. Unfortunately, LLM fact-checkers also significantly change trust in news when they are wrong or provide inconclusive answers. Our results demonstrate both the potential for LLMs to correct false information at scale but also their potential to taint the truth at scale.}
\end{abstract}

\section{Introduction}
Both experts and the public are concerned about the acceptance and dissemination of online misinformation \cite{ecker2024misinformation}. Consequently, researchers and practitioners have implemented multiple interventions to limit the impact of false information. One of the most widely deployed and well-studied interventions is warning labels from fact-checkers. Fact-checker warning labels have been shown to reduce people\change{'s} belief in, trust in, and sharing intentions of misinformation \cite{martel2023misinformation}.  

Given today’s fragmented and polarized media environment \cite{flamino2023political}, one of the most surprising and promising results from the literature on fact checking and corrections is how consistently effective they are across \textit{partisan differences}. Most studies find little to no evidence of political ideology significantly impacting the effectiveness of corrections (e.g., \citet{horne2019rating}, \citet{pennycook2020implied}, \citet{porter2022political}, \citet{horne2025does}), even though there is evidence that baseline truth discernment differs across Republicans and Democrats \cite{clemm2023truth, dobbs2023democrats}. Partisanship can sometimes influence correction strength, but people across partisan groups generally update their beliefs in the correct direction. For example, \citet{hameleers2020misinformation} showed that fact-checkers can “lower agreement with attitudinally congruent political misinformation” and \citet{weeks2015emotions} demonstrated that “exposure to corrections improves belief accuracy, regardless of emotion or partisanship”. Furthermore, backfire effects from corrections are uncommon \cite{wood2019elusive, guess2020does}, and even those who distrust fact-checkers are influenced by fact checks \cite{martel2024fact}.  

While fact checking has proven to be a highly effective intervention, it is also highly resource-intensive and requires human labor to evaluate claims at scale and in real time. Over the past decade, researchers have designed and evaluated a variety of methods to automate corrections and reduce the resources needed to deploy corrections at scale \cite{zhou2020survey,  dierickx2023automated}. The success of these automated methods has largely been limited, often due to an inability to generalize far outside the model training data \cite{horne2023ethical}. However, recent Large Language Models (LLMs), which are trained on massive datasets and can augment that training data through web search, have shown an impressive ability to generalize across a wide range of tasks. Studies have shown that LLMs perform relatively well in fact-checking and misinformation detection tasks \cite{ zhou2024correcting, kuznetsova2025generative, huang2025unmasking}, and in related tasks like rating the credibility of news outlets \cite{yang2025accuracy}.

Along with these advancements, social media platforms have increasingly moved away from human third-party fact-checkers, due in part to political pressure in the U.S. \cite{horne2025despite}. Instead, general-use LLMs have been embedded on these platforms. For example, on the social media platform X/Twitter, Grok and Perplexity are LLM chatbot accounts that reply to users who mention them. While there are still ``ethical and societal limitations'' of many current LLM application implementations \cite{narayanan2025search}, people are using them for fact-checking. According to preliminary work by \citet{renault2025grok}, there is ``growing bi-partisan demand for LLM fact-checking integrated into social media platforms'' and \change{the launch of the Grok and Perplexity chatbots is associated with a lower use of Community Notes}, a once popular system that enables users to attach fact-check notes to content \cite{allen2022birds}. 

Importantly, LLM-based fact-checkers – whether they are explicitly built for fact-checking tasks or not - differ from traditional fact-checking in ways that may affect how users respond to corrections. Of particular interest to our study, LLMs can be configured to be ideologically aligned with their users via the political content emphasized in their responses, the sources cited, and the configured persona \cite{kroger2025don}. The most prominent example of an ideologically configured LLM is the chatbot Grok, which has been explicitly configured to be “not woke” and provide politically conservative responses \cite{Zeff2023Grok, Melimopoulos2025Grok}. Preliminary work suggests that Republican X/Twitter users tend to select Grok over the more neutral Perplexity for fact-checks \cite{renault2025grok}. As a result, corrections from LLM fact-checkers may be influenced by partisan preferences. This raises critical questions about whether the abovementioned effectiveness of fact-checking across partisan groups extends to corrections delivered by politically congruent or incongruent LLMs. Hence, \textbf{the goal of this paper is to systematically test the effectiveness of fact-checking across partisan differences in the new context of ideologically aligned LLMs.} \change{Specifically, we ask two research questions:} 

\begin{quote}
\change{\textbf{RQ1:} How effective is fact-checking information from politically congruent and incongruent LLM chatbots at (a) reducing trust in false information and (b) increasing trust in true information?}
\end{quote}

\begin{quote}
\change{\textbf{RQ2:} How does the political congruency of the news headlines being fact-checked impact the effectiveness of fact-checking information from LLM chatbots?}
\end{quote}

Using data from two within-subjects experiments (n=705), where politically diverse Americans interact with real-time LLM chatbots, we show that the robust effectiveness of fact-checking largely survives the transition from human fact-checkers to politically aligned LLMs. This effectiveness largely holds regardless of the LLM's political congruency with the user. \change{That is, trust in false information decreases and trust in true information increases after interacting with both politically congruent and incongruent chatbots. However, perceived chatbot congruency plays a small role when headlines are politically distant. Specifically, trust in correctly labeled true headlines increases less when politically incongruent chatbots check distant headlines and increases more when congruent or neutral chatbots check distant headlines. This pattern was not found for false headlines.} Perhaps of more concern, our experiment also demonstrates that people update veracity judgments based on LLM fact-checks even when those checks are incorrect or inconclusive. \change{We further discuss these results, their implications, and avenues for future work in this paper.} Supplemental materials for this paper can be found here: \url{https://osf.io/fyjcx/overview?view_only=a23d51fb25bb4a74bf84548a4a9c6bd7}

\section{Related Work}

\subsection{\change{Fact-checking Interventions}}
Exposure to corrections, fact-checks, or content warning labels typically reduces belief in, trust in, accuracy perception of, and sharing intention of misinformation \cite{martel2023misinformation}, even though those effects are sometimes modest and the influence of misperceptions may not be eliminated in the long-term \cite{grady2021nevertheless}. Evidence suggests that people are particularly susceptible to misinformation that reinforces their existing political positions \cite{frenda2013false, xiong2023effects}. However, as discussed in the introduction to this paper, content warning labels are generally effective across a variety of individual differences, including partisanship. This is not to say individual differences have no effect. For example, evidence from survey experiment data suggest that labels from Facebook and Twitter (prior to Elon Musk's purchase of the platform) were less effective on Republicans \cite{lees2022twitter, jennings2023asymmetric}. Other manipulations, such as manipulating the source of a correction, can interact with these individual differences. For example, \citet{horne2025does} found that warning labels from generic AI (not LLMs) were significantly more effective at reducing trust than labels from other sources for people who reported a low trust in news organizations, and \citet{yaqub2020effects} found that labels from fact-checkers \change{reduced the} false news sharing intent of Democrats more than Republicans. Still, people across partisan groups generally updated their trust and sharing intention in the correct direction.

The design of fact-checking interventions can play a role, albeit a small one, in their effectiveness. Most notably, adding explanations or reasoning can improve effectiveness. For example, adding more details to a fact-check \cite{brashier2021timing, kreps2022covid} or providing details about how warning labels are generated \cite{horne2019rating, epstein2022explanations} can increase effectiveness.

People may follow corrections even when they are imperfect or incorrect. Studies have found that erroneous warning labels can decrease trust in accurate content (tainted truth effect) \cite{freeze2021fake, horne2025people}. On the other hand, other work has found that people may interpret false content that is left unlabeled as true (implied truth effect) \cite{pennycook2020implied}. Hence, both labeling coverage and accuracy are important in fact-checking systems.

\subsection{\change{LLM Fact-checkers}}
Just as LLMs represent a ``significant shift from traditional search engines'' \cite{narayanan2025search}, they also represent a significant shift from traditional fact-checking. We identify two primary reasons why LLM-based fact checkers should be studied in their own right rather than treated as substitutes for human fact checkers. First, to the extent that fact checkers are perceived as acting on behalf of news consumers, LLMs possess fundamentally different agentic properties than human agents. Second, LLMs alter the information-processing conditions under which veracity judgments are formed.

We conceptualize the relationship between a news consumer and a fact checker as an agentic relationship in which the principal delegates the task of assessing news veracity to an agent \cite{jensen2019theory}. Whereas human agents are typically modeled as utility-maximizing, goal-pursuing individuals \cite{eisenhardt1989agency}, LLM outputs are shaped by training data, model architecture, and design constraints rather than intrinsic preferences or incentives. This delegation creates an adverse selection problem: news consumers often lack information about an LLM’s training, domain competence, or potential biases, limiting their ability to assess agent quality ex ante. However, because LLMs can provide explanations, evidence, and interactive follow-up on demand, principals may rely less on ex ante quality signals and more on ex post output characteristics when evaluating the agent’s suitability. Thus, LLM-based fact checking reshapes the traditional agency relationship and merits further investigation.

To explain how news consumers process fact-checking outputs under these new conditions, we further draw on the Heuristic–Systematic Model (HSM; \citet{chaiken1980heuristic}). HSM posits that individuals rely either on heuristic cues or on systematic processing, depending on effort considerations and confidence needs \cite{chaiken2012theory}. Importantly, what constitutes ``effort'' in HSM is not fixed but depends on the information environment. Traditional fact-checking by human experts often presents users with only a simple heuristic cue (e.g., true/false, verified/unverified). Systematic processing in this environment may require significantly more effort on behalf of the user, such as reading further outside of the platform where the content was encountered. Indeed, studies on HSM in the context of human fact-checking suggests that users often treat warning labels as heuristic cues, rather than mechanisms to initiate systematic processing or elaboration \cite{koch2023effects, horne2025people}. In contrast, LLM-based fact checking can easily provide explanations, evidence, and interactive follow-ups, substantially increasing opportunities for elaboration at a reduced effort (even if that evidence is incorrect). LLMs therefore can impact the choice between heuristic and systematic processing, ultimately changing the reliance on their advice. 

Studies about human interaction with LLM chatbots used for fact-checking are nascent, but so far, their use in interventions has produced mixed results. LLMs have been shown to be persuasive in tasks like reducing conspiracy beliefs \cite{costello2024durably}, reducing science skepticism \cite{hornsey2025using}, persuading voters \cite{lin2025persuading}, \change{and purchasing behavior \cite{salvi2026commercialpersuasionaimediatedconversations}}. On the other hand, through a between-subjects experiment, \citet{deverna2024fact} demonstrated that fact-checking information from ChatGPT, a general-use LLM, does ``not necessarily enhance users’ abilities to discern headline accuracy or promote accurate news sharing''. Moreover, the persuasiveness of LLMs can have negative side effects. For example, a recent preliminary study found that LLMs were ``as effective at increasing conspiracy belief as decreasing it'' \cite{costello2026large}. Building on the above literature, we next describe the experimental design and analytical approach used to examine how politically congruent and incongruent LLM fact-checkers influence trust updating.

\begin{figure*}[ht!]
    \centering
    \subfloat[\centering Example of Left Bot]{{{\frame{\includegraphics[trim=0 0cm 0cm 0,clip,width=5.5cm]{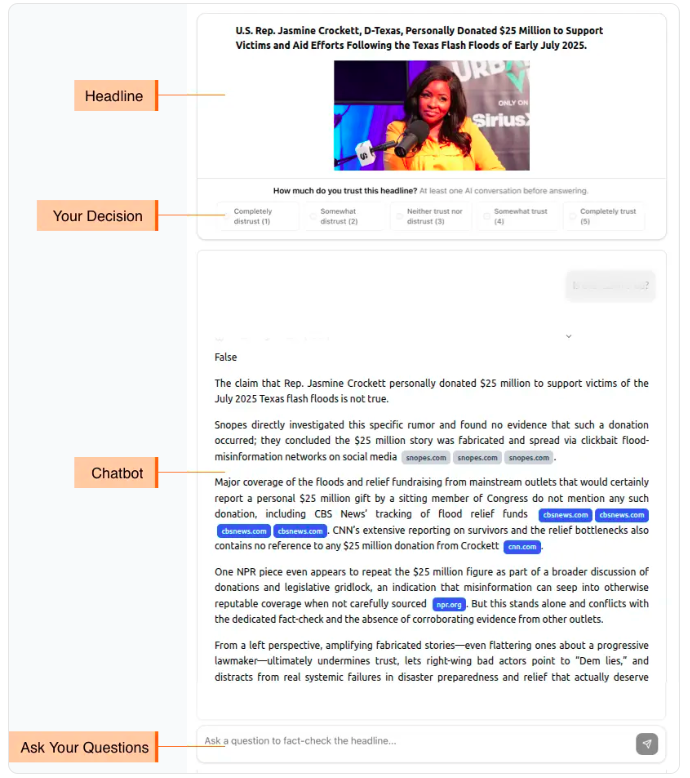} }}}}\hfill
    \subfloat[\centering Experiment Flows]{{{\includegraphics[trim=0 0cm 0cm 0,clip,width=11cm]{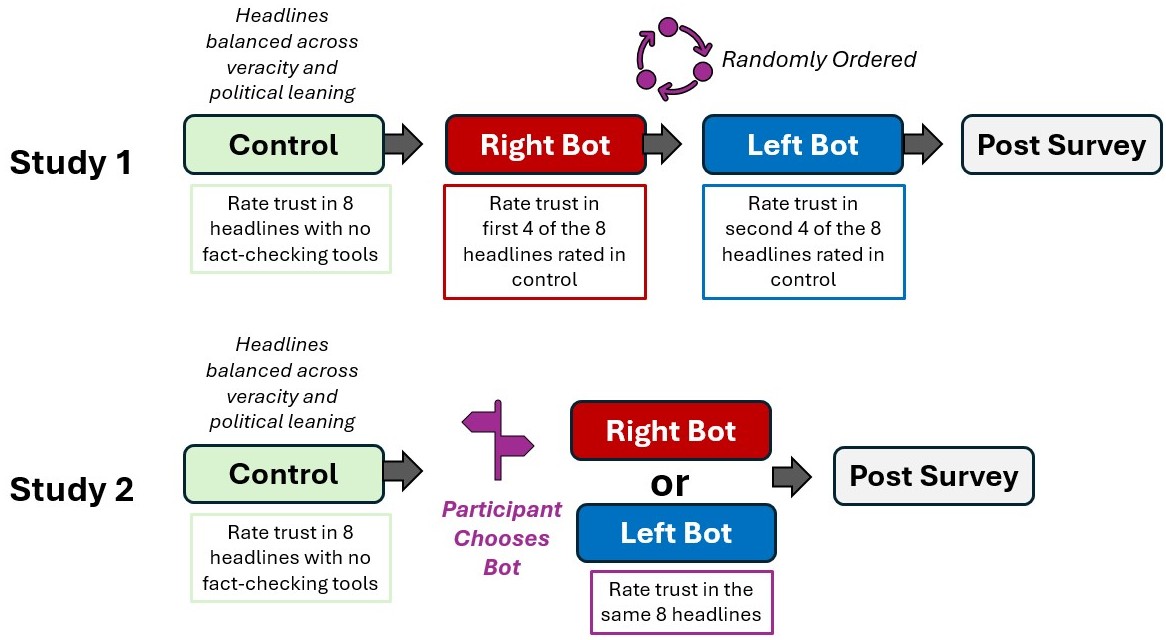} }}}\\
    \caption{(a) An example of the left bot interface. (b) The experiment flows for both study 1 and study 2. Both studies used within-subjects designs, where participants receive both the control condition and the treatment conditions. In study 1, participants use both bots in a random order. Importantly, they are told they will use two different bots and are told when moving from one to another. In study 2, participant choose which bot to use.}%
    \label{fig:screenshots}%
\end{figure*}

\section{Methods}

\subsection{LLM Chatbot Design and Interface}

The fact-checking chatbot used in this study was a web-based conversational AI system. The system was designed to assist participants in evaluating news headlines’ veracity through fact-checking capabilities with real-time search, where participants could ask unlimited questions about a given headline. The chatbot used GPT 5.1 with medium reasoning effort and was integrated with the Exa API to perform real-time web searches as a LLM tool. Web searches returned up to 25 results per query, and the LLM could submit multiple queries in response to a single prompt. Those web results had to be published after January 1, 2025 and from the United States.  Search results from Exa API included LLM-generated article summaries, publication dates, and source URLs.

The chatbot supported our experimental conditions via media source filtering and system prompt framing. Together, these manipulations produced chatbot responses that exhibited political bias through the 
\textit{information presented} (source-level bias) and \textit{its framing} (language-level bias). Specifically, the chatbot was configured to restrict search results to specific news domains based on political leaning. The news domains that each chatbot was restricted to can be found in the supplemental materials. \change{These domains were grouped into left and right political leanings using AllSides Media Bias Chart\footnote{\url{https://www.allsides.com/media-bias/media-bias-chart}}, which has been used similarly in research on media bias (e.g., \citet{dai2025media}).} Further, the chatbot's system prompt was configured to reflect different political perspectives, instructing the model to convey principles typically associated with either left-leaning or right-leaning viewpoints. The prompt framing was adapted from existing empirical studies on political bias of LLMs~\cite{agiza2024politune, ChenHYSL24, smith2025investigating}. \change{Specifically, we adopted the ideological persona assignment template from \citet{ChenHYSL24}, explicitly instructing the model to reflect a left- or right-leaning ideology and convey its associated principles so that participants could infer its stance from the response.}

To ensure the LLM features were consistent across experimental conditions, the chatbot followed a structured response protocol of: (1) provide an initial verdict label (True, False, or Unverifiable), (2) deliver concise responses under 300 words, (3) cite at least three sources from the configured media group, (4) include one to two fact-checking sources (i.e., Snopes, Reuters Fact Check) if possible, and (5) pose follow-up questions to encourage continued engagement. A complete system prompt can be found in the supplement.
\change{To verify that the chatbot adhered to this protocol, we evaluated first responses delivered to participants across both studies, and report compliance rates for each instruction.
The compliance rates were high overall: 99.6\% of responses stayed under 300 words, 100\% began with a verdict label, 95.6\% contained at least one markdown citation, 93.8\% met the three-source requirement when applicable, and 97.3\% closed with a follow-up question. Detailed results are provided in the supplement.}

The interface presented headlines in a card format displaying the headline text, an associated image, and a 5-point trust rating scale ranging from ``completely distrust'' to ``completely trust.'' Below the headline, a chat panel allowed participants to interact with the chatbot through a text input field (see Figure~\ref{fig:screenshots}a). To promote natural conversational interaction and ensure meaningful prompt data, the headline text was automatically appended as contextual data to participants' first prompt before being sent to the LLM. Participants were instructed to formulate questions without restating the headline. Chatbot responses were streamed in real-time, with a collapsible ``Reasoning Process'' panel displaying ongoing search queries and sources being consulted. Inline citations appeared as color-coded badges indicating the political orientation of each source (blue for left-leaning, red for right-leaning), with hover cards providing article titles, publication dates, and content summaries. Participants were required to complete at least one conversational exchange before submitting their trust rating for each headline.

\subsection{Experimental Setup}
\textbf{Study 1:} We recruited 412 U.S. adults from the survey platform Prolific to participate in our within-subjects experiment. These participants came from two samples: (a) 294 participants were from U.S. representative sample across age, gender, and political affiliation, and (b) 118 participants were balanced across U.S. political affiliation (Republicans, Independents, and Democrats). There were 150 participants, balanced across U.S. political affiliation, recruited for three rounds of pilot studies. \change{We chose to balance these samples across political leaning to ensure sufficient variation and statistical power for analyses involving political congruency.} We made significant changes to the design after the first pilot and those responses were removed from the dataset. The remaining responses were added to our study sample of 300 participants from a U.S. representative sample (as no changes were made between pilot \#2 and the full experiment). Eight respondents were removed due to low quality responses as measured by multiple attention checks. 


Each participant was given a random set of eight political news headlines, balanced across veracity and political leaning. For each headline they were asked how much they trust the information in the headline on a 5-point scale (\textit{completely distrust} to \textit{completely trust}). After rating trust in each headline, all participants were given two LLM chatbots to fact-check those same headlines. They were first given either a chatbot designed to be right-leaning or left-leaning. The participants then used the chatbot to fact check the first four of the eight headlines they initially rated in control. After at least one conversation with the chatbot, participants again rated their trust in the headline. After fact-checking and rerating trust in four headlines, participants were then given the chatbot of the opposite political leaning to fact check the last four of the eight headlines they initially rated in control. While participants were required to complete at least one round of conversation with the chatbot for each headline, they were free to complete as many rounds of conversation as they wanted before submitting their new trust rating. A flow chart can be found in Figure \ref{fig:screenshots}b.

Prior to being assigned a chatbot, participants were told they would use two different chatbots for fact-checking. They were also told when the chatbots switched. For each chatbot, they were shown a small tutorial of the interface. In this tutorial, they were not explicitly told that the bots were configured to be liberal or conservative but instead were given a list of example news outlets that each LLM would use to fact check. For the left bot, these example news outlets were The Guardian, The Huffington Post, MSNBC, The Daily Beast, and Mother Jones. For the right bot, these example news outlets were The American Conservative, Breitbart, Fox News, The Daily Caller, and Newsmax. \change{The order in which these chatbots were shown to the participants was randomized to mitigate order effects, and subsequent analysis showed that the order in which a participant saw the chatbots did not impact our dependent variable. This model can be found in the supplement.}

After completing the task across both chatbots, all participants were redirected to a survey to capture participants' political leaning and perceptions of each chatbot.

\noindent \textbf{Study 2:} We recruited 293 adults from a U.S. representative sample across age, gender, and political affiliation from the survey platform Prolific. Seven respondents were removed due to low quality responses as measured by multiple attention checks. The experimental setup of study 2 mimicked that of study 1, except participants chose which chatbot to use and used the chosen chatbot to fact check all eight headlines. In the choice menu, participants were given the same instructions and details about each chatbots design via example news outlets. The choice menu and instructions can be in the supplement. In both studies, participants were paid \$4 for 20 minutes. In both studies, the median time of participation was slightly under 20 minutes. \change{The Exa API cost was approximately \$0.48 per participant, on a rate of \$7 per 1{,}000 search requests and \$1 per 1{,}000 summaries, with a maximum of three searches per conversational round. The cost of the GPT-5 API was not significant, given the rate of \$2.50 per 1M input tokens and \$15.00 per 1M output tokens.}

\subsection{Headline Selection}
To ensure that headlines were recent and balanced across political congruency, we follow a two step process. First, we created a set of 200 political news headlines, half false. Following a similar process done in \citet{horne2025does}, we randomly selected political headlines that were published in the last six months. False headlines were selected from third-party fact-checking sites, such as Snopes, PolitiFact, FactCheck.org, or Reuters Fact Check. True headlines were selected from both recently fact-checked articles and articles from Reuters and the Associated Press. Less than half of the true headlines were from fact-checking organizations \change{(38\%)}, while the rest were from these two reputable news outlets \change{(62\%)}. This approach allows our true headlines to be represented by a mixture of typical factual content and true headlines that required fact-checking. From the selected articles, we used the headline and an image in the article to create the stimuli.

Second, we launched a pre-test in which a separate sample of 356 participants indicated ``\textit{how good for the Democrats versus Republicans the headline would be if it was true}'' on a 5-point Likert scale from \textit{very favorable} to \textit{very unfavorable} (similar to the approach used by \citet{chen2023makes}). \change{This sample of participants was mutually exclusive of the 705 participants in the main study, and they were recruited from a U.S. representative sample stratified across age, gender, and political affiliation on Prolific.} Participants rated 20 randomly selected headlines from the 200 headlines described above. With these ratings, we computed the average political leaning of each headline, where very favorable to Democrats was 5 and very unfavorable to Democrats was 1. We filtered out 43 headlines that had an average rating of neither favorable nor unfavorable (ratings between 2.75 and 3.25), leaving us with 157 headlines. Lastly, to create a balanced set of headlines across veracity and political leaning, we randomly sampled 33 headlines from each of the four groups: true right, true left, false right, and false left. The final set of headlines used can be found in the supplement.

\subsection{Data Analysis}
Since our data captured repeated measures across participants, our primary analysis was done using mixed effects regression with random intercepts for participants. For ease of interpretation and to examine the impact of LLM correctness, we built models separately for true headlines and false headlines. In each model, our dependent variable was trust change (\textbf{Trust $\Delta$}), the difference between a participant's trust in control and in treatment for the same headline. A positive Trust $\Delta$ means trust in the headline increased, while a negative Trust $\Delta$ means trust in the headline decreased. Given our interest was in the impact of political congruency on effectiveness, our independent variables included:
\begin{itemize}
    \item Chatbot headline verdict (\textbf{Bot said \textit{X}}) - Each chatbot was required to start its answer with a verdict. This verdict could either be \textbf{True}, \textbf{False}, or \textbf{Unverifiable}. We used this verdict as a categorical variable in our model.
    \item Perceived political distance (or incongruence) from chatbots (\textbf{Bot $\neq$}) - We computed the perceived political distance by taking the absolute value of the difference between participants' political leaning (\textbf{PL}) and their perceived political leaning of the chatbot (\textbf{PBL}). Both were measured on 7-point scales. For example, if a participant was `slightly liberal' and perceived the chatbot as `moderate', then their bot distance was 1.
    \begin{itemize}[leftmargin=1.5cm, itemsep=5pt, parsep=0pt]
    \item[\textbf{(EQ1)}] \change{Bot $\neq$ $=$ $|$PL $-$ PBL$|$}
    \end{itemize}
    \item Political \change{distance} with headlines
    (\textbf{Headline \change{$\neq$}}) - We computed each headline's political \change{distance} with participants by taking the absolute value of the difference between the headline political leaning (\textbf{HPL}) and the participants' political leaning (\textbf{PL}). Headline political leaning scores were transformed from 5-point to 7-point scales to match the scale of participants' leaning.
    \begin{itemize}[leftmargin=1.5cm, itemsep=5pt, parsep=0pt]
    \item[\textbf{(EQ2)}] \change{Headline $\neq$ $=$ $|$PL $-$ $\left(\frac{6}{4}\right)\,$HPL$ + 1$$|$}
    \end{itemize} 

    \item We also examined the interaction between headline \change{distance} and perceived chatbot \change{distance}. Hence, our primary regression model was: 
    \begin{itemize}[leftmargin=1.5cm, itemsep=5pt, parsep=0pt]
    \item[\textbf{(EQ3)}] Trust $\Delta$ $=$ Bot said True $+$ Bot said False $+$ Bot $\neq$ $+$ Headline $\change{\neq}$ $+$ Bot $\neq$ $*$ Headline $\change{\neq}$
\end{itemize}
\end{itemize}

The same within-subjects model setup was used independently for both study 1 and study 2. We also constructed two between-subjects models using data from both studies to examine the impact of choosing a fact-checking bot rather than being randomly assigned one. \change{These models can be found in the supplemental materials.} \change{In addition to the participant-only random effect model, we fit mixed-effects regression models with random intercepts for both participants and headlines to account for heterogeneity across headlines. These models produced the same results as the models with only random intercepts for participants and can be found in the supplemental materials.} Lastly, we used Kolmogorov–Smirnov test (KS) and Cliff's Delta (Cliffs $\delta$) to further describe the data. 

\section{Study Descriptives}
To contextualize our results, we provide descriptive statistics of the participants and chatbots \change{in Table \ref{tab:demo}.} 

\begin{table}[h]
\centering
\begin{tabular}{lcc}
\toprule
 & \textbf{Study 1} & \textbf{Study 2} \\
\midrule
\textbf{Average age} & 45.7 & 45.3 \\\midrule
Identified as \textbf{Women} & 51.0\% & 51.0\% \\
Identified as \textbf{White} & 70.6\% & 75.6\% \\
Identified as \textbf{Conservative} & 36.7\% & 36.2\% \\
\bottomrule
\end{tabular}
\caption{\change{Participant demographics.}}
\label{tab:demo}
\end{table}


\begin{figure}[ht!]
    \centering
    \subfloat[\centering Study 1 Manipulation]{{{\includegraphics[trim=0 0cm 0cm 0,clip,width=4cm]{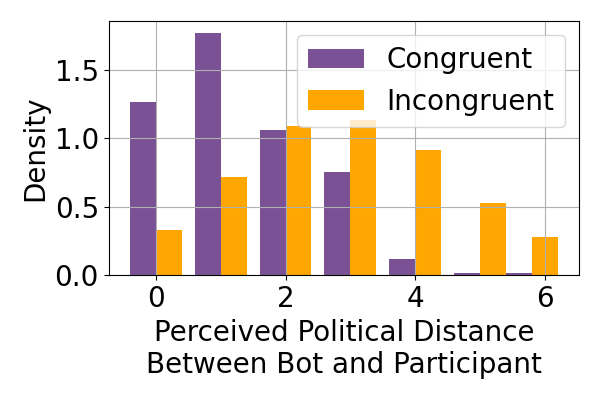} }}}
    \subfloat[\centering Study 1 vs. Study 2]{{{\includegraphics[trim=0 0cm 0cm 0,clip,width=4cm]{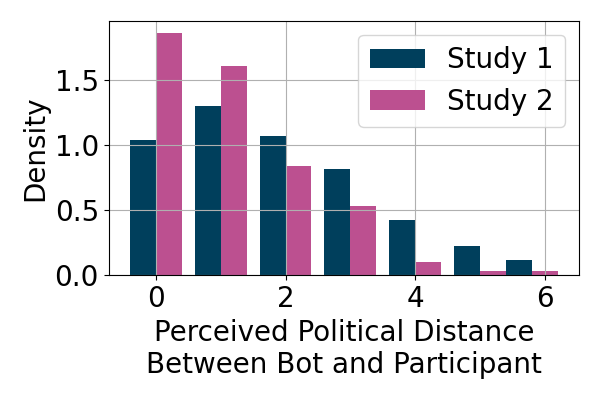} }}}
    \caption{(a) Histograms of perceived political distance between bot and participants across congruent and incongruent bots. This plot excludes politically moderate participants. These distributions significantly differ with a medium effect size (KS = 0.225, p = $2.4e-61$, Cliff's $\delta$ = 0.29) (b) Histograms of perceived political distance between bot and participants across all participants in study 1 and study 2. These distributions significantly differ with a large effect size (KS = 0.396, p $= 9.26e-96$, Cliff's $\delta$ = 0.54).}
    \label{fig:botdists}%
\end{figure}




\noindent \textbf{Manipulation Check:}
In Figure \ref{fig:botdists}a, we show the distributions of the participants' perceived political distance from the bots across congruent and incongruent bots. A \textit{congruent} bot is when the participant is the same political leaning as the bot was designed to be  (e.g., right bot, conservative participant). \textit{Incongruent} is when the participant is the opposite political leaning of what the bot was designed to be (e.g., right bot, liberal participant). Overall, most people correctly perceived the right/left bot manipulation, and the perceived political distance was significantly smaller when participants interacted with the congruent bot. We provide additional manipulation check results in the supplement.

\noindent \textbf{Study 2 Participant Choices:}
As described above, study 2 allowed participants to select which bot they would use for fact-checking. We found that bot choice largely reflected the political leaning of participants, with 88.8\% of liberals selecting the left bot, 64.2\% of conservatives selecting the right bot, and 75.4\% of moderates selecting the left bot. This reflects prior work that showed fact checkers were most likely to be selected when they confirmed prior attitudes \cite{hameleers2020misinformation}. We show the full distribution of 7-point participant political leaning to bot choice in the supplemental materials. 

\noindent \textbf{Bot Correctness:}
The accuracy of each LLM fact-checker was in-part dependent on the web sources each was restricted to. If possible, the LLM had to include at least one fact-checking source, which helped the chatbot make correct verdicts. During each study, both the left bot and right bot correctly labeled the majority of headlines. The right bot was slightly less accurate than the left bot in both studies. The left bot labeled false headlines correctly between 90\% and 91\% of the time and labeled true headlines correctly between 88\% and 91\% of the time. The right bot labeled false headlines correctly between 85\% and 91\% of the time and labeled true headlines correctly between 80\% and 86\% of the time. We provide the full distributions of correctness in the supplemental materials. We also tested the accuracy of each LLM across both headlines and user sessions. This analysis showed that the LLMs were highly consistent in their verdict across interactions with the same headline. Further, we tested each LLM's accuracy on headlines prior to the study with a static user prompt and found similar results. These results can also be found in the supplement.

\section{Results}

\begin{figure}[ht!]
    \centering
    \subfloat[\centering True Headlines - Study 1]{{{\includegraphics[trim=0 0cm 0cm 0,clip,width=4cm]{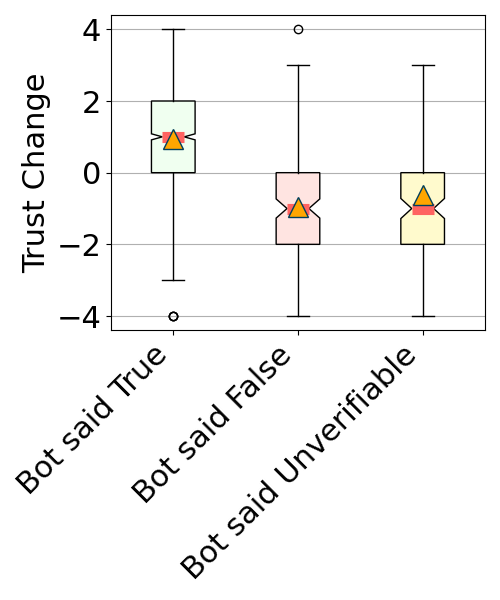} }}}
    \subfloat[\centering False Headlines - Study 1]{{{\includegraphics[trim=0 0cm 0cm 0,clip,width=4cm]{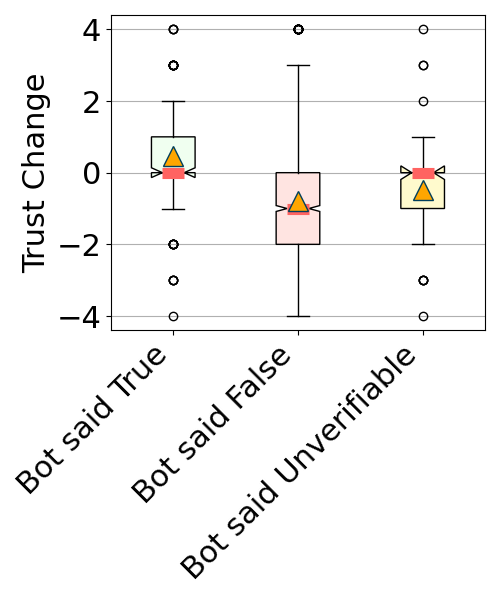} }}}\\
    \subfloat[\centering True Headlines - Study 2]{{{\includegraphics[trim=0 0cm 0cm 0,clip,width=4cm]{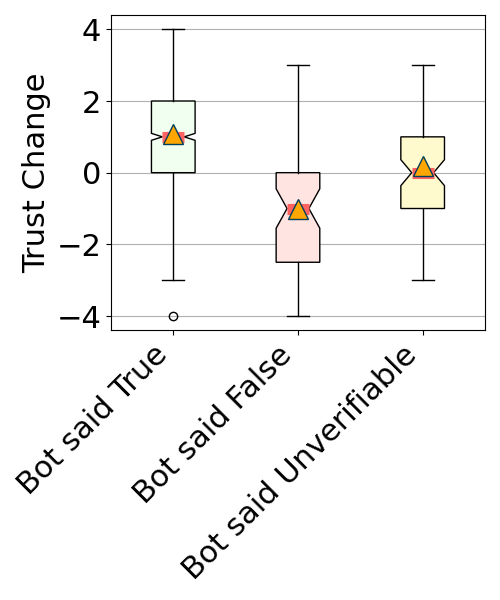} }}}
    \subfloat[\centering False Headlines - Study 2]{{{\includegraphics[trim=0 0cm 0cm 0,clip,width=4cm]{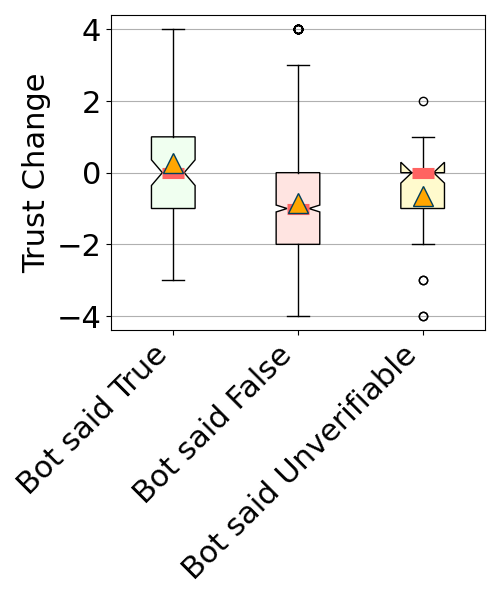} }}}\\
    \caption{Distributions of trust change by LLM verdict for true and false headlines in study 1 (a, b) and study 2 (c, d).}%
    \label{fig:boxes}%
\end{figure}

\textbf{LLM fact-checkers are generally effective at shifting trust in news.} As can be seen in panels \ref{fig:boxes}a and \ref{fig:boxes}c, when the bot correctly identified a true headline, trust went up by an average of 0.93 in study 1 and 1.09 in study 2. As can be seen in panels \ref{fig:boxes}b and \ref{fig:boxes}d, when the bot correctly identified a false headline, trust decreased by an average of -0.78 in study 1 and -0.85 in study 2. These changes in trust were statistically significant.  When comparing the distribution of trust in control with the distribution of trust in the treatment, correct verdicts changed trust across both true and false headlines with medium to large effect sizes (see Table \ref{tab:ks_cliff_combined}).

\begin{table}[ht!]
\centering
\begin{tabular}{lccc}
\toprule
\multicolumn{4}{c}{\textbf{(a) Correct verdicts}} \\
\midrule
 & \textbf{KS} & \textbf{Sig} & \textbf{Cliff's} $\delta$ \\
\midrule
\textbf{Study 1 true}  & 0.368 & $p = 6.64e-84$ & -0.47 \\
\textbf{Study 1 false} & 0.343 & $p = 3.26e-76$ & 0.37 \\
\textbf{Study 2 true}  & 0.423 & $p = 4.72e-84$ & -0.54 \\
\textbf{Study 2 false} & 0.391 & $p = 6.04e-73$ & 0.41 \\
\midrule
\multicolumn{4}{c}{\textbf{(b) Incorrect verdicts}} \\
\midrule
 & \textbf{KS} & \textbf{Sig} & \textbf{Cliff's} $\delta$ \\
\midrule
\textbf{Study 1 true}  & 0.370 & $p = 8.91e-09$ & 0.37 \\
\textbf{Study 1 false} & 0.242 & $p = 0.000815$ & -0.19 \\
\textbf{Study 2 true}  & 0.490 & $p = 6.33e-06$ & 0.42 \\
\textbf{Study 2 false} & 0.221 & $p = 0.0466$ & -0.14 \\
\bottomrule
\end{tabular}
\caption{\change{Trust change statistics (Kolmogorov-Smirnov) and effect sizes for (a) correct and (b) incorrect verdicts.}}
\label{tab:ks_cliff_combined}
\end{table}

\begin{figure}[h]
    \centering
    \subfloat[\centering True Headlines - Study 1]{{\includegraphics[trim=0 0cm 0cm 0,clip,width=4.1cm]{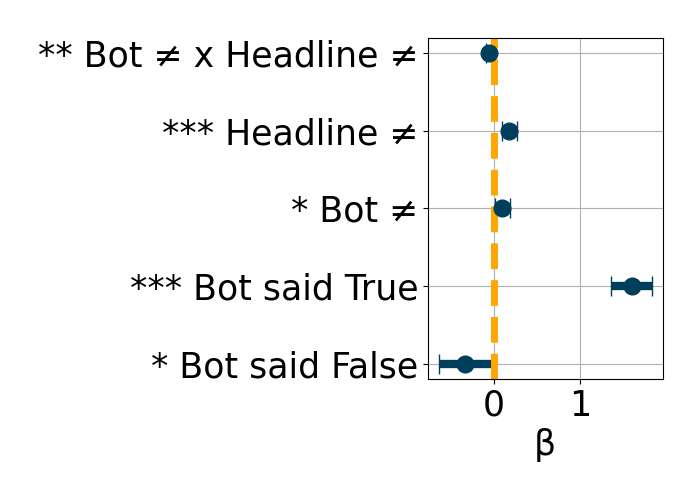} }}
    \subfloat[\centering False Headlines - Study 1]{{{\includegraphics[trim=0 0cm 0cm 0,clip,width=4.1cm]{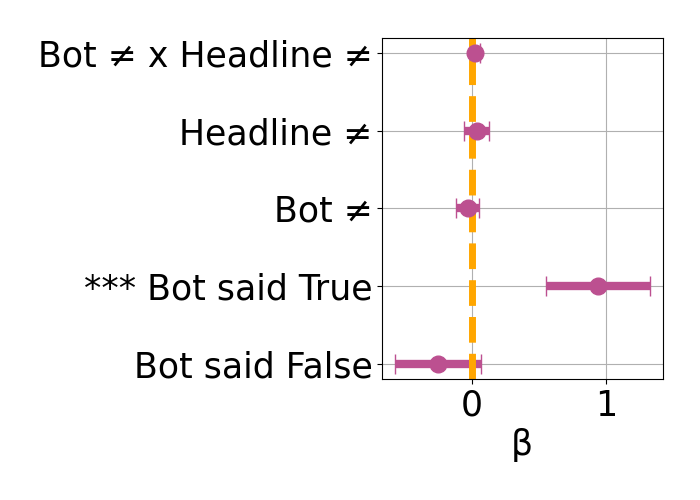} }}}\\
    \subfloat[\centering True Headlines - Study 2]{{\includegraphics[trim=0 0cm 0cm 0,clip,width=4.1cm]{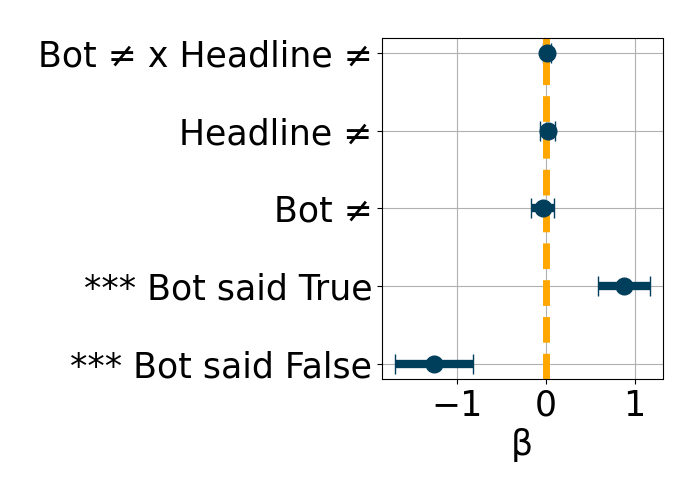} }}
    \subfloat[\centering False Headlines - Study 2]{{{\includegraphics[trim=0 0cm 0cm 0,clip,width=4.1cm]{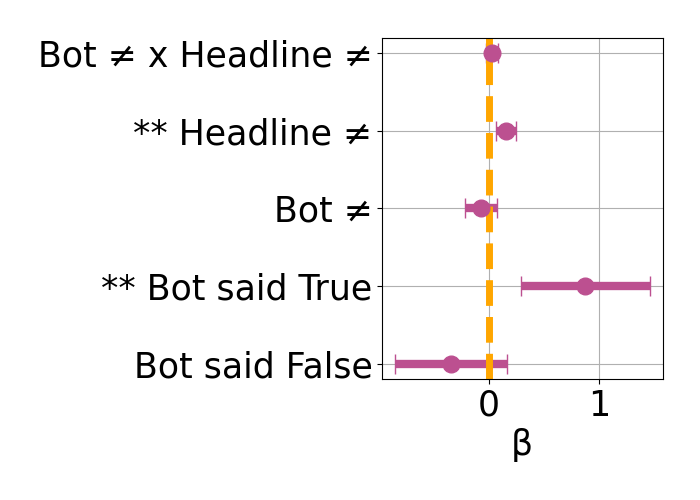} }}}\\
    \caption{Coefficient plots from four mixed effects models with random intercepts for participants where the dependent variable is trust $\Delta$ and the independent variables are the bots fact-check verdict, the perceived political distance from the bot (Bot $\neq$), the political congruency of the headline (Headline $\change{\neq}$), and their interaction. We show separate models for true headlines (a, c) and false headlines (b, d) (EQ3). Statistical significance is indicated by asterisks in each plot, where *** $p < 0.001$, ** $p < 0.01$, * $p < 0.05$.}%
    \label{fig:mixedeff}%
\end{figure}

\textbf{LLM fact-checkers change trust in news even when they are wrong or provide inconclusive answers.}
Trust also significantly shifted when bots provided incorrect answers (`True' instead of `False', vice versa). As shown in panels \ref{fig:boxes}a and \ref{fig:boxes}c, when the bot incorrectly identified a true headline, trust decreased by -0.96 on average in study 1 and -1.02 in study 2. As shown in panels \ref{fig:boxes}b and \ref{fig:boxes}d, when the bot incorrectly identified a false headline, trust increased by +0.45 on average in study 1 and +0.27 in study 2. When comparing trust distributions across control and treatment, incorrect verdicts changed trust with a small to medium effect size, no matter the headline ground truth (see Table \ref{tab:ks_cliff_combined}.)

Trust also changed when the bots provided inconclusive answers (`Unverifiable'), but the direction and magnitude of change were inconsistent across studies. When the bot provided an `Unverifiable' verdict, trust in true headlines decreased by -0.63 on average in study 1 but increased by +0.19 in study 2. For false headlines the direction of change was consistent, decreasing by -0.49  on average in study 1 and -0.66 in study 2. In general, inconclusive answers decreased trust in headlines, no matter their veracity, although this shift was not always significant.

\begin{figure}[h!]
    \centering
    \subfloat[\centering Trust $\Delta$ = Bot $\neq$ x Headline $\change{\neq}$]{{{\includegraphics[trim=0 0cm 0cm 0,clip,width=7.2cm]{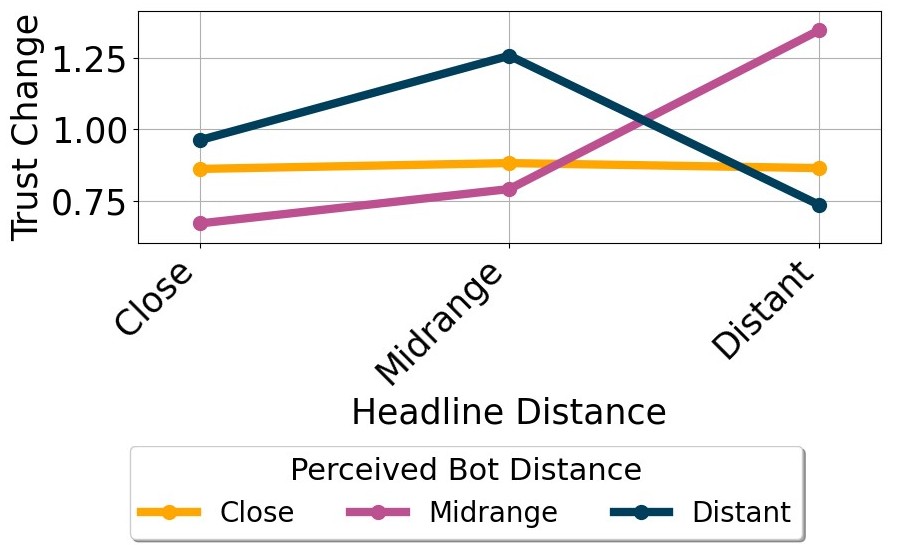} }}}\\
    \subfloat[\centering Only True Close Headlines (Left side of (a))]{{{\includegraphics[trim=0 0cm 0cm 0,clip,width=4.1cm]{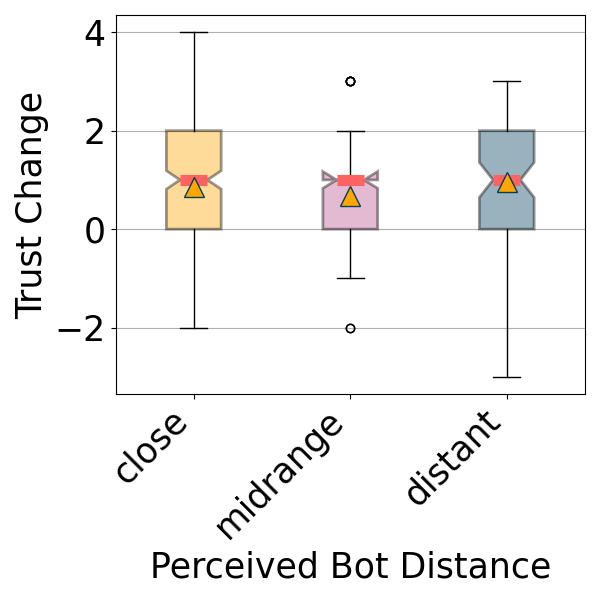} }}}
    \subfloat[\centering Only True Distant Headlines (Right side of (a))]{{{\includegraphics[trim=0 0cm 0cm 0,clip,width=4.1cm]{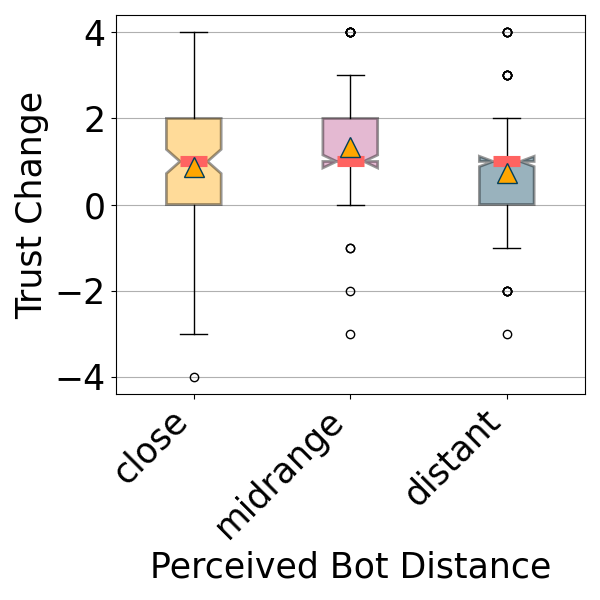} }}}\\
    \caption{\change{(Top) Interaction between perceived bot congruency and headline congruency for correctly labeled true headlines in study 1. In this plot, both perceived bot distance and headline distance were binned into three groups via tertiles. (Bottom) Trust change distributions across perceived bot distance for (b) correctly labeled true politically close headlines and (c) correctly labeled true politically distant headlines. The distribution of trust change only significantly differed between midrange bots and distant bots for distant headlines. No other trust change distributions significantly differed (despite the visual difference of midrange headlines in (a), trust change distributions did not significantly differ across bot distance for more moderate headlines).}}%
    \label{fig:interactions}%
\end{figure}

\begin{figure*}[ht!]
    \centering
    \subfloat[\centering Study 1: Trust Before Treatment]{{{\includegraphics[trim=0 0cm 0cm 0,clip,width=7.5cm]{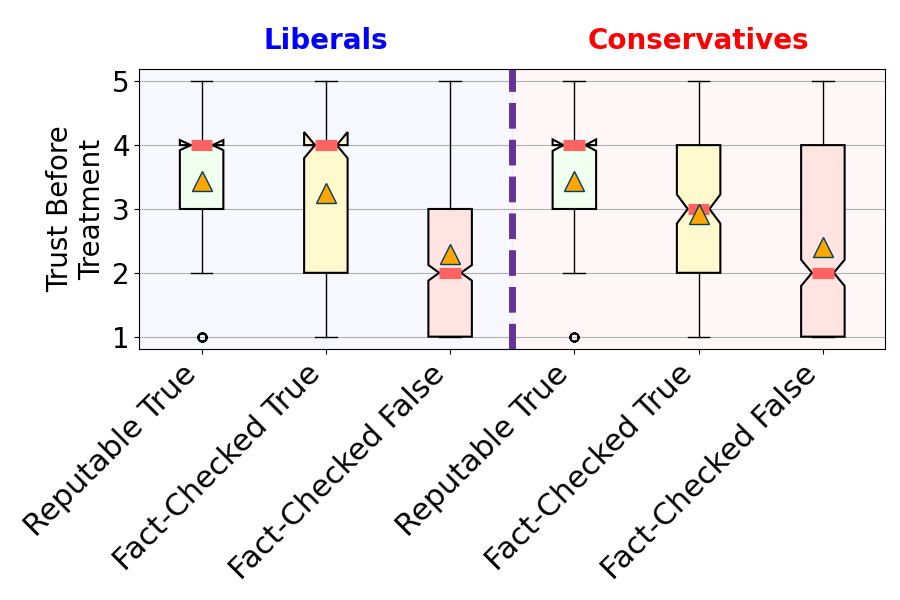} }}}
    \subfloat[\centering Study 1: Trust After Treatment]{{{\includegraphics[trim=0 0cm 0cm 0,clip,width=7.5cm]{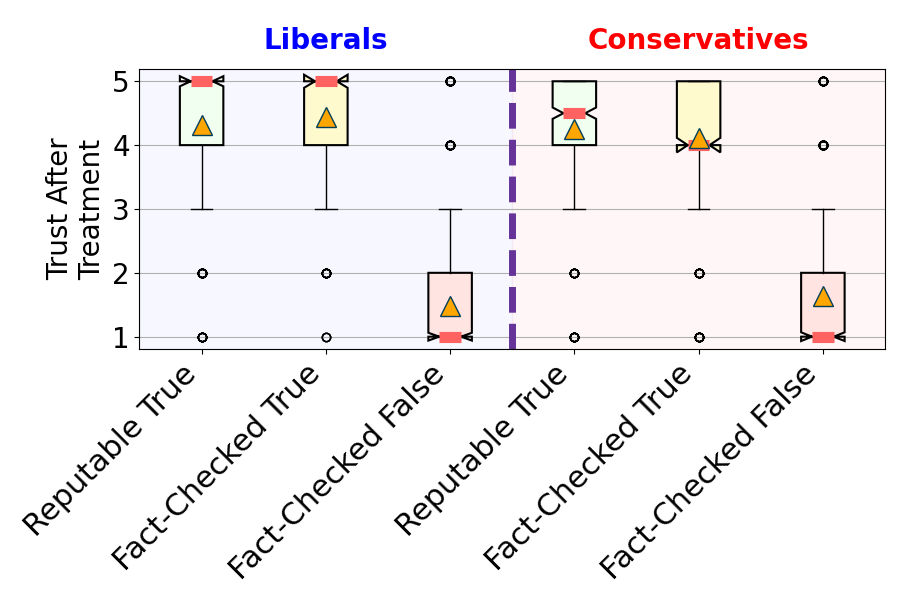} }}}\\
    \subfloat[\centering Study 2: Trust Before Treatment]{{{\includegraphics[trim=0 0cm 0cm 0,clip,width=7.5cm]{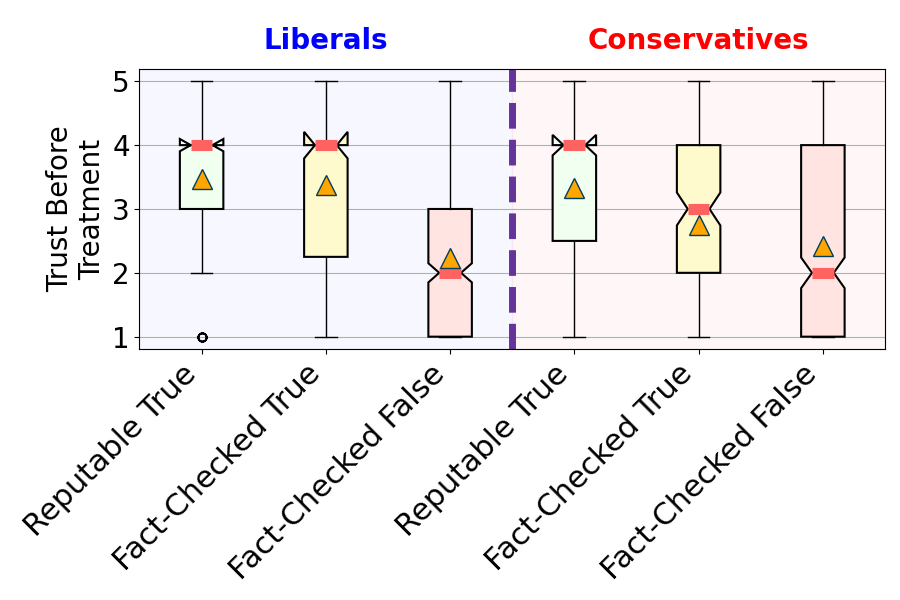} }}}
    \subfloat[\centering Study 2: Trust After Treatment]{{{\includegraphics[trim=0 0cm 0cm 0,clip,width=7.5cm]{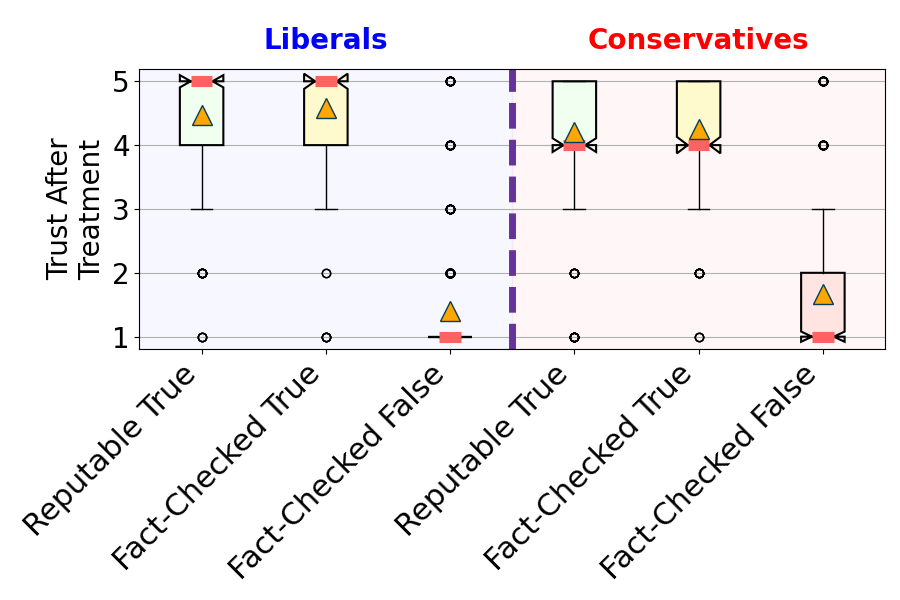} }}}\\
    \caption{\change{Distributions of trust in true headlines that came from reputable sources, true headlines that were fact-checked, and false headlines across conservatives and liberals in study 1 and study 2. In (a) and (c), we show the distributions of initial trust, while in (b) and (d) we show the distributions of trust after correct treatment (bot labeled true headline as true, labeled false headlines as false). These plots ignore the political congruency of bots and headlines.}}%
    \label{fig:true_headline_types_poli}%
\end{figure*}

\textbf{\change{Perceived political congruency between the participant and the bot matters only when headlines are politically distant.}} In Figure \ref{fig:mixedeff}, we show coefficient plots from our mixed effects models for effectiveness (EQ3) across study 1 and study 2. We found a significant relationship between the perceived political distance from the bot (Bot $\neq$) and trust change for true headlines in study 1 (Figure \ref{fig:mixedeff}a, $\beta$ = 0.099, $p < 0.05$). This was not found for false headlines.


We further explored the impact of bot political congruency in study 1 by creating distributions of \textit{designed congruency} (i.e., a conservative participant interacting with the right-leaning bot is congruent, regardless of their perception of the bot) and \textit{perceived congruency} (i.e., a bot was congruent if the participant perceived the bot to be the same political leaning as them). \change{We show plots of these distributions in the supplemental materials.} For both metrics, trust in true headlines increased significantly more when the bot was politically congruent but, these effects were negligible, \change{reflecting the mixed-effects regression results} (Designed congruency: KS = 0.117, p $= 0.000176$, Cliff's $\delta$ = 0.11; Perceived congruency: KS = 0.092, p $= 0.0396$, Cliff's $\delta$ = 0.09).

\change{Perceived bot congruency played a larger role in trust change when headlines were politically distant.} Specifically, in study 1, perceived distance from the bot significantly interacted with headline distance (Figure \ref{fig:mixedeff}a, Bot $\neq$ x Headline $\neq$, $\beta$ = -0.051, $p < 0.01$). As shown in Figure \ref{fig:interactions}a, \change{trust in correctly labeled true headlines increased less when distant bots checked distant headlines and increased more when moderate bots checked distant headlines.}

\change{To explain this result, we plotted the distributions of trust change across perceived bot distance for politically close true headlines (Figure \ref{fig:interactions}b) and politically distant true headlines (Figure \ref{fig:interactions}c). In Figure \ref{fig:interactions}b, none of the distributions significantly differed from each other according to a KS test. In Figure \ref{fig:interactions}c, the distribution of trust change for perceived midrange bots significantly differed from the distribution of trust change for perceived distant bots (KS = 0.235, p $< 0.001$, Cliff's $\delta$ = 0.27). This result suggests that the interaction shown in Figure \ref{fig:interactions}a is one-sided. That is, the magnitude of trust change significantly differed between distant bots checking distant headlines and midrange bots checking distant headlines (right side of Figure \ref{fig:interactions}a). But, trust change did not significantly differ across close, midrange, and distant bots when the headlines being checked were politically closer (middle and left side of Figure \ref{fig:interactions}a). Again, these patterns were not found for false headlines.} 

In study 2, headline congruency alone had a significant impact on trust change for false headlines, where trust decreased less for congruent false headlines (Figure \ref{fig:mixedeff}d, $\beta$ = 0.157, $p < 0.01$). \change{There was not a significant interaction between perceived bot congruency and headline congruency in study 2, but this is likely due to the fact that most participants chose politically congruent bots.}

\change{\textbf{LLM fact-checkers are effective across different types of headlines.}} \change{In general, participants initially trusted true headlines that had to be fact-checked less than they trusted true headlines from reputable sources, although these differences were small to negligible. Further, participants had some ability to discern true and false information on average, initially trusting false headlines less than true headlines. This result aligns with prior studies that have demonstrated that people are able to discern false news across topics \cite{pfander2025spotting}, and interventions simply improve upon this baseline discernment \cite{martel2023misinformation, horne2025people}.}



\change{In Figure \ref{fig:true_headline_types_poli}, we show the distributions of trust before and after treatment for true headlines that came from reputable sources, true headlines that had to be fact-checked, and false headlines. We break these distributions across conservatives and liberals. Conservatives initially trusted true headlines that had to be fact-checked less than liberals did (study 1: KS = 0.131, $p < 0.05$, Cliff's $\delta$ = 0.15; study 2: KS = 0.213, $p = 0.01$, Cliff's $\delta$ = 0.28), but there were no significant differences between conservatives and liberals initial trust in true headlines from reputable sources nor their initial trust in false headlines.}

\change{While both conservatives and liberals were helped by interacting with the bots, there were differences in headline trust after treatment. In study 1, conservatives still trusted true headlines that had to be fact-checked less than liberals (KS = 0.149, $p < 0.05$, Cliff's $\delta$ = 0.16). In study 2, where participants could choose their bot, conservatives trusted both true headlines from reputable sources and those that had to be fact-checked less than liberals (Reputable True: KS = 0.168, $p < 0.01$, Cliff's $\delta$ = 0.17; Fact-Checked True: KS = 0.220, $p < 0.001$, Cliff's $\delta$ = 0.23).} 


\change{These small differences across partisanship may be partially explained by conservatives distrust of fact-checking websites like Snopes \cite{robertson2020uses}, which had to be included in the bot's explanation if possible.} \change{Still, despite these differences, LLM fact-checkers that correctly labeled headlines were effective across all three types of headlines and across both liberals and conservatives, increasing trust in true news and decreasing trust in false news.}


\section{Discussion}
In this study, we examined how Americans' trust in true and false political headlines changed after having conversations with ideologically configured LLM chatbots. We showed that fact-checking remains effective at changing trust when performed by LLMs instead of humans. We found that this effect largely held regardless of the chatbot’s political alignment with the user \change{or the political alignment of the news.}

\change{One concern with traditional fact-checking interventions is that they are focused on decreasing trust in false information rather than increasing trust in true information. Prior work suggests that people are generally skeptical about online news \cite{fletcher2019generalised}, and those who distrust reputable news media may turn to less reliable alternative outlets \cite{hameleers2022whom}. Prior work also suggests that people are ``better able to rate false news as inaccurate than to rate true news as accurate'' \cite{pfander2025spotting}, further illustrating the need for interventions that increase trust in true news. Our results suggest that conversational LLMs can effectively serve both roles, increasing trust in true information and decreasing trust in false information.}

\change{Surprisingly, we found that the perceived political congruency of chatbots played only a weak role in intervention effectiveness. Despite participants often correctly perceiving our ideological manipulations, in most cases, participants' trust in headlines changed no matter the political congruency of the chatbot. While we found some partisan differences in headline trust after treatment, these differences were small. These results may reflect the tendency of people to view AI as objective and accurate (often called \textit{Machine Heuristics}; \citet{sundar2020rise}). Prior work has shown that even simple AI source cues can reduce perceptions of partisan bias \cite{cloudy2023str} and increase the effectiveness of content labels \cite{horne2025does}. Our results suggests that conversational agents can be effective at decreasing trust in false news across partisanship. Our results also suggest that conversational agents can be effective at increasing trust in true news across partisanship, albeit to a lesser extent.} 

\change{These positive results are limited by conversational agents ability to accurately fact-check news, as our results also showed that headline trust changed when the chatbot was wrong. This result is perhaps unsurprising given that prior work demonstrated that people listen to content warning labels when they are wrong \cite{freeze2021fake, horne2025does}. Further, prior work has demonstrated that LLMs generate responses that ``sound certain'', even when they are wrong \cite{rathi2025humans, narayanan2025search}. While this finding may be unsurprising, it is particularly concerning because not all LLM applications are configured for the task of fact-checking.} Our study used LLM chatbots that were customized and carefully configured for the task of fact-checking. While we restricted what news outlets they could use to manipulate political leaning, we required both chatbots to also use third-party, human-driven fact-checkers like Snopes. This setup ensured the chatbots correctly labeled headlines the vast majority of the time. General-use chatbots or chatbots configured for other purposes may not be correct as often. For example, \citet{deverna2024fact} showed that ChatGPT only labeled true headlines as \textit{true} 15\% of the time. 

\change{The persuasive power of LLM chabots combined with the importance of configuring LLM applications for specific tasks highlights a key concern:} \textit{who} has power to configure and deploy these models could ``gain disproportionate influence over public discourse'' \cite{kroger2025don}. Due to the resources needed to build and host LLMs, only a few companies control the ownership of them \cite{sathish2024llempower}, meaning that the power of steering LLMs towards specific ideological positions, framings, and ``truths'' is also only in the hands of a few. \change{Accountability for and safeguards against mistakes made by public-facing LLMs are issues that current regulatory frameworks are not well-equipped for, particularly in the United States \cite{fang2024large}. Hence, academic research should continue to both build safeguards for LLM chatbots as well as educate the public about them.}

\change{\textbf{Contributions:} To summarize, this paper makes three key contributions. First, it extends the current research on LLM fact-checking to ideologically configured chatbots and uses a real conversational environment, reflecting those already embedded on social media platforms. Second, this work thoroughly examines intervention effectiveness across a balanced set of politically congruent and incongruent headlines, ensuring validity across partisanship. Third, it advances the conversation about both the positive and negative implications of automating online fact-checking.}

\textbf{Limitations and Future Work:}
Overall, the results of this work are both similar and notably different to the prior work on LLM fact-checkers. Both our work and the prior found that wrong verdicts form LLMs can significantly influence users beliefs \cite{deverna2024fact}. At the same time, the prior work found that LLM chatbots did \textit{not} significantly help participants’ ability to discern headline accuracy. We suspect that our results are different from the results of \citet{deverna2024fact} for three reason: (1) the prior work used ChatGPT, a general use LLM without configuration for the task of fact-checking or real-time access to web search, (2) the recency of news headlines used in the prior work ranged widely, while our headlines were all published/fact-checked within 6 months of the experiment, and (3) the prior work used screenshots of chatbot output, while our study used a real-time chatbot with unlimited user interactions. This comparison is not to say the results from \citet{deverna2024fact} are irrelevant but that they were found under different conditions. While our experimental setup was designed for realism, it is also limited by lack of control. There are many unknown variables stemming from the dynamically generated text, real-time online information, and diverse user prompts. Future work should analyze how specific combinations of user prompts and generated text influenced fact-checking effectiveness. \change{Furthermore, future work should examine how methodological choices (real chatbot versus screenshots) in conversational LLM experiments impact results.}

Given the conditions under which our results were found, it is likely that LLMs currently embedded on social media platforms also significantly alter trust in information - both correctly and incorrectly. This result underscores the potential of LLMs to help correct misinformation at scale, but this result also highlights the need to design safeguards and better understand how information seekers interact with LLMs. \change{Future work can move towards better designs by studying how individuals reason about and verify information from generated by conversational LLMs.}
  
\section{Conclusion}
Using data from two within-subjects experiments, this paper examined how conversations with ideologically-configured LLM chatbots influence trust in true and false political news. We found that LLM fact-checkers shifted trust in both true and false headlines, regardless of the chatbot’s political alignment with the user. \change{Perceived political congruency between the participant and the bot matters only when headlines were politically distant: trust increased less when checked by ideologically distant chatbots and more when checked by more ideologically moderate ones. Importantly, LLM fact-checkers also influenced trust when they provided incorrect or inconclusive information}, highlighting both their potential to correct inaccurate content at scale and their potential to taint the truth at scale.

\bibliography{scibib}

\begin{thebibliography}{64}
\providecommand{\natexlab}[1]{#1}

\bibitem[{Agiza, Mostagir, and Reda(2024)}]{agiza2024politune}
Agiza, A.; Mostagir, M.; and Reda, S. 2024.
\newblock Politune: Analyzing the impact of data selection and fine-tuning on economic and political biases in large language models.
\newblock In \emph{Proceedings of the AAAI/ACM Conference on AI, Ethics, and Society}, volume~7, 2--12.

\bibitem[{Allen, Martel, and Rand(2022)}]{allen2022birds}
Allen, J.; Martel, C.; and Rand, D.~G. 2022.
\newblock Birds of a feather don’t fact-check each other: Partisanship and the evaluation of news in Twitter’s Birdwatch crowdsourced fact-checking program.
\newblock In \emph{Proceedings of the 2022 CHI conference on human factors in computing systems}, 1--19.

\bibitem[{Brashier et~al.(2021)Brashier, Pennycook, Berinsky, and Rand}]{brashier2021timing}
Brashier, N.~M.; Pennycook, G.; Berinsky, A.~J.; and Rand, D.~G. 2021.
\newblock Timing matters when correcting fake news.
\newblock \emph{Proceedings of the National Academy of Sciences}, 118(5): e2020043118.

\bibitem[{Chaiken(1980)}]{chaiken1980heuristic}
Chaiken, S. 1980.
\newblock Heuristic versus systematic information processing and the use of source versus message cues in persuasion.
\newblock \emph{Journal of personality and social psychology}, 39(5): 752.

\bibitem[{Chaiken and Ledgerwood(2012)}]{chaiken2012theory}
Chaiken, S.; and Ledgerwood, A. 2012.
\newblock A theory of heuristic and systematic information processing.
\newblock \emph{Handbook of theories of social psychology}, 1: 246--266.

\bibitem[{Chen et~al.(2024)Chen, He, Yan, Shi, and Lerman}]{ChenHYSL24}
Chen, K.; He, Z.; Yan, J.; Shi, T.; and Lerman, K. 2024.
\newblock How Susceptible are Large Language Models to Ideological Manipulation?
\newblock In Al{-}Onaizan, Y.; Bansal, M.; and Chen, Y., eds., \emph{Proceedings of the 2024 Conference on Empirical Methods in Natural Language Processing, {EMNLP} 2024, Miami, FL, USA, November 12-16, 2024}, 17140--17161. Association for Computational Linguistics.

\bibitem[{Chen, Pennycook, and Rand(2023)}]{chen2023makes}
Chen, X.; Pennycook, G.; and Rand, D. 2023.
\newblock What makes news sharable on social media?
\newblock \emph{Journal of Quantitative Description: Digital Media}, 3.

\bibitem[{Clemm~von Hohenberg(2023)}]{clemm2023truth}
Clemm~von Hohenberg, B. 2023.
\newblock Truth and bias, left and right: testing ideological asymmetries with a realistic news supply.
\newblock \emph{Public Opinion Quarterly}, 87(2): 267--292.

\bibitem[{Cloudy, Banks, and Bowman(2023)}]{cloudy2023str}
Cloudy, J.; Banks, J.; and Bowman, N.~D. 2023.
\newblock The str (AI) ght scoop: Artificial intelligence cues reduce perceptions of hostile media bias.
\newblock \emph{Digital Journalism}, 11(9): 1577--1596.

\bibitem[{Costello et~al.(2026)Costello, Pelrine, Kowal, Arechar, Godbout, Gleave, Rand, and Pennycook}]{costello2026large}
Costello, T.~H.; Pelrine, K.; Kowal, M.; Arechar, A.~A.; Godbout, J.-F.; Gleave, A.; Rand, D.; and Pennycook, G. 2026.
\newblock Large language models can effectively convince people to believe conspiracies.
\newblock \emph{arXiv preprint arXiv:2601.05050}.

\bibitem[{Costello, Pennycook, and Rand(2024)}]{costello2024durably}
Costello, T.~H.; Pennycook, G.; and Rand, D.~G. 2024.
\newblock Durably reducing conspiracy beliefs through dialogues with AI.
\newblock \emph{Science}, 385(6714): eadq1814.

\bibitem[{Dai et~al.(2025)Dai, Cao, Wang, Pang, Xu, Ng, and Chua}]{dai2025media}
Dai, S.; Cao, Z.; Wang, W.; Pang, L.; Xu, J.; Ng, S.~K.; and Chua, T.-S. 2025.
\newblock Media source matters more than content: Unveiling political bias in llm-generated citations.
\newblock In \emph{Proceedings of the 2025 Conference on Empirical Methods in Natural Language Processing}, 17267--17287.

\bibitem[{DeVerna et~al.(2024)DeVerna, Yan, Yang, and Menczer}]{deverna2024fact}
DeVerna, M.~R.; Yan, H.~Y.; Yang, K.-C.; and Menczer, F. 2024.
\newblock Fact-checking information from large language models can decrease headline discernment.
\newblock \emph{Proceedings of the National Academy of Sciences}, 121(50): e2322823121.

\bibitem[{Dierickx, Lind{\'e}n, and Opdahl(2023)}]{dierickx2023automated}
Dierickx, L.; Lind{\'e}n, C.-G.; and Opdahl, A.~L. 2023.
\newblock Automated fact-checking to support professional practices: systematic literature review and meta-analysis.
\newblock \emph{International Journal of Communication}, 17: 21.

\bibitem[{Dobbs et~al.(2023)Dobbs, DeGutis, Morales, Joseph, and Swire-Thompson}]{dobbs2023democrats}
Dobbs, M.; DeGutis, J.; Morales, J.; Joseph, K.; and Swire-Thompson, B. 2023.
\newblock Democrats are better than Republicans at discerning true and false news but do not have better metacognitive awareness.
\newblock \emph{Communications Psychology}, 1(1): 46.

\bibitem[{Ecker et~al.(2024)Ecker, Tay, Roozenbeek, Van Der~Linden, Cook, Oreskes, and Lewandowsky}]{ecker2024misinformation}
Ecker, U.~K.; Tay, L.~Q.; Roozenbeek, J.; Van Der~Linden, S.; Cook, J.; Oreskes, N.; and Lewandowsky, S. 2024.
\newblock Why misinformation must not be ignored.
\newblock \emph{American Psychologist}.

\bibitem[{Eisenhardt(1989)}]{eisenhardt1989agency}
Eisenhardt, K.~M. 1989.
\newblock Agency theory: An assessment and review.
\newblock \emph{Academy of management review}, 14(1): 57--74.

\bibitem[{Epstein et~al.(2022)Epstein, Foppiani, Hilgard, Sharma, Glassman, and Rand}]{epstein2022explanations}
Epstein, Z.; Foppiani, N.; Hilgard, S.; Sharma, S.; Glassman, E.; and Rand, D. 2022.
\newblock Do explanations increase the effectiveness of AI-crowd generated fake news warnings?
\newblock In \emph{Proceedings of the International AAAI Conference on Web and Social Media}, volume~16, 183--193.

\bibitem[{Fang and Perkins(2024)}]{fang2024large}
Fang, A.; and Perkins, J. 2024.
\newblock Large language models (LLMs): Risks and policy implications.
\newblock \emph{MIT Sci. Policy Rev.}, 5(2024): 134--45.

\bibitem[{Flamino et~al.(2023)Flamino, Galeazzi, Feldman, Macy, Cross, Zhou, Serafino, Bovet, Makse, and Szymanski}]{flamino2023political}
Flamino, J.; Galeazzi, A.; Feldman, S.; Macy, M.~W.; Cross, B.; Zhou, Z.; Serafino, M.; Bovet, A.; Makse, H.~A.; and Szymanski, B.~K. 2023.
\newblock Political polarization of news media and influencers on Twitter in the 2016 and 2020 US presidential elections.
\newblock \emph{Nature Human Behaviour}, 7(6): 904--916.

\bibitem[{Fletcher and Nielsen(2019)}]{fletcher2019generalised}
Fletcher, R.; and Nielsen, R.~K. 2019.
\newblock Generalised scepticism: How people navigate news on social media.
\newblock \emph{Information, communication \& society}, 22(12): 1751--1769.

\bibitem[{Freeze et~al.(2021)Freeze, Baumgartner, Bruno, Gunderson, Olin, Ross, and Szafran}]{freeze2021fake}
Freeze, M.; Baumgartner, M.; Bruno, P.; Gunderson, J.~R.; Olin, J.; Ross, M.~Q.; and Szafran, J. 2021.
\newblock Fake claims of fake news: Political misinformation, warnings, and the tainted truth effect.
\newblock \emph{Political behavior}, 43(4): 1433--1465.

\bibitem[{Frenda et~al.(2013)Frenda, Knowles, Saletan, and Loftus}]{frenda2013false}
Frenda, S.~J.; Knowles, E.~D.; Saletan, W.; and Loftus, E.~F. 2013.
\newblock False memories of fabricated political events.
\newblock \emph{Journal of Experimental Social Psychology}, 49(2): 280--286.

\bibitem[{Grady, Ditto, and Loftus(2021)}]{grady2021nevertheless}
Grady, R.~H.; Ditto, P.~H.; and Loftus, E.~F. 2021.
\newblock Nevertheless, partisanship persisted: Fake news warnings help briefly, but bias returns with time.
\newblock \emph{Cognitive research: principles and implications}, 6: 1--16.

\bibitem[{Guess and Coppock(2020)}]{guess2020does}
Guess, A.; and Coppock, A. 2020.
\newblock Does counter-attitudinal information cause backlash? Results from three large survey experiments.
\newblock \emph{British Journal of Political Science}, 50(4): 1497--1515.

\bibitem[{Hameleers, Brosius, and de~Vreese(2022)}]{hameleers2022whom}
Hameleers, M.; Brosius, A.; and de~Vreese, C.~H. 2022.
\newblock Whom to trust? Media exposure patterns of citizens with perceptions of misinformation and disinformation related to the news media.
\newblock \emph{European journal of communication}, 37(3): 237--268.

\bibitem[{Hameleers and Van~der Meer(2020)}]{hameleers2020misinformation}
Hameleers, M.; and Van~der Meer, T.~G. 2020.
\newblock Misinformation and polarization in a high-choice media environment: How effective are political fact-checkers?
\newblock \emph{Communication research}, 47(2): 227--250.

\bibitem[{Horne(2025)}]{horne2025does}
Horne, B.~D. 2025.
\newblock Does the Source of a Warning Matter? Examining the Effectiveness of Veracity Warning Labels Across Warners.
\newblock In \emph{Proceedings of the International AAAI Conference on Web and Social Media}, volume~19, 823--836.

\bibitem[{Horne and Craig(2025)}]{horne2025despite}
Horne, B.~D.; and Craig, M.~J. 2025.
\newblock Despite Meta Ending Its Third-Party Fact-Checking Program, Most People Still Want Fact-Checkers on Social Media.

\bibitem[{Horne and Nevo(2025)}]{horne2025people}
Horne, B.~D.; and Nevo, D. 2025.
\newblock People adhere to content warning labels even when they are wrong due to ecologically rational adaptations.
\newblock \emph{Scientific Reports}, 15(1): 1--13.

\bibitem[{Horne et~al.(2019)Horne, Nevo, O'Donovan, Cho, and Adal{\i}}]{horne2019rating}
Horne, B.~D.; Nevo, D.; O'Donovan, J.; Cho, J.-H.; and Adal{\i}, S. 2019.
\newblock Rating Reliability and Bias in News Articles: Does AI Assistance Help Everyone?
\newblock In \emph{Proceedings of the International AAAI Conference on Web and Social Media}, volume~13, 247--256.

\bibitem[{Horne, Nevo, and Smith(2023)}]{horne2023ethical}
Horne, B.~D.; Nevo, D.; and Smith, S.~L. 2023.
\newblock Ethical and safety considerations in automated fake news detection.
\newblock \emph{Behaviour \& Information Technology}, 1--22.

\bibitem[{Hornsey et~al.(2025)Hornsey, Smith, Pearson, Bretter, and Nylund}]{hornsey2025using}
Hornsey, M.~J.; Smith, A.~E.; Pearson, S.; Bretter, C.; and Nylund, J.~L. 2025.
\newblock Using conversational AI to reduce science skepticism.
\newblock \emph{Current Opinion in Psychology}, 102216.

\bibitem[{Huang et~al.(2025)Huang, Yi, Yu, and Xu}]{huang2025unmasking}
Huang, T.; Yi, J.; Yu, P.; and Xu, X. 2025.
\newblock Unmasking digital falsehoods: A comparative analysis of LLM-based misinformation detection strategies.
\newblock In \emph{2025 8th International Conference on Advanced Algorithms and Control Engineering (ICAACE)}, 2470--2476. IEEE.

\bibitem[{Jennings and Stroud(2023)}]{jennings2023asymmetric}
Jennings, J.; and Stroud, N.~J. 2023.
\newblock Asymmetric adjustment: Partisanship and correcting misinformation on Facebook.
\newblock \emph{New Media \& Society}, 25(7): 1501--1521.

\bibitem[{Jensen and Meckling(1976)}]{jensen2019theory}
Jensen, M.~C.; and Meckling, W.~H. 1976.
\newblock Theory of the firm: Managerial behavior, agency costs and ownership structure.
\newblock In \emph{Corporate governance}, 77--132. Gower.

\bibitem[{Koch, Frischlich, and Lermer(2023)}]{koch2023effects}
Koch, T.~K.; Frischlich, L.; and Lermer, E. 2023.
\newblock Effects of fact-checking warning labels and social endorsement cues on climate change fake news credibility and engagement on social media.
\newblock \emph{Journal of Applied Social Psychology}, 53(6): 495--507.

\bibitem[{Kreps and Kriner(2022)}]{kreps2022covid}
Kreps, S.~E.; and Kriner, D.~L. 2022.
\newblock The COVID-19 infodemic and the efficacy of interventions intended to reduce misinformation.
\newblock \emph{Public Opinion Quarterly}, 86(1): 162--175.

\bibitem[{Kr{\"o}ger and Barkett(2025)}]{kroger2025don}
Kr{\"o}ger, P.; and Barkett, E. 2025.
\newblock Don't Change My View: Ideological Bias Auditing in Large Language Models.
\newblock \emph{arXiv preprint arXiv:2509.12652}.

\bibitem[{Kuznetsova et~al.(2025)Kuznetsova, Makhortykh, Vziatysheva, Stolze, Baghumyan, and Urman}]{kuznetsova2025generative}
Kuznetsova, E.; Makhortykh, M.; Vziatysheva, V.; Stolze, M.; Baghumyan, A.; and Urman, A. 2025.
\newblock In generative AI we trust: can chatbots effectively verify political information?
\newblock \emph{Journal of Computational Social Science}, 8(1): 15.

\bibitem[{Lees, McCarter, and Sarno(2022)}]{lees2022twitter}
Lees, J.; McCarter, A.; and Sarno, D.~M. 2022.
\newblock Twitter’s disputed tags may be ineffective at reducing belief in fake news and only reduce intentions to share fake news among Democrats and Independents.
\newblock \emph{Journal of Online Trust and Safety}, 1(3).

\bibitem[{Lin et~al.(2025)Lin, Czarnek, Lewis, White, Berinsky, Costello, Pennycook, and Rand}]{lin2025persuading}
Lin, H.; Czarnek, G.; Lewis, B.; White, J.~P.; Berinsky, A.~J.; Costello, T.; Pennycook, G.; and Rand, D.~G. 2025.
\newblock Persuading voters using human--artificial intelligence dialogues.
\newblock \emph{Nature}, 1--8.

\bibitem[{Martel and Rand(2023)}]{martel2023misinformation}
Martel, C.; and Rand, D.~G. 2023.
\newblock Misinformation warning labels are widely effective: A review of warning effects and their moderating features.
\newblock \emph{Current Opinion in Psychology}, 101710.

\bibitem[{Martel and Rand(2024)}]{martel2024fact}
Martel, C.; and Rand, D.~G. 2024.
\newblock Fact-checker warning labels are effective even for those who distrust fact-checkers.
\newblock \emph{Nature Human Behaviour}, 8(10): 1957--1967.

\bibitem[{Melimopoulos(2025)}]{Melimopoulos2025Grok}
Melimopoulos, E. 2025.
\newblock What is Grok and why has Elon Musk’s chatbot been accused of anti-Semitism?
\newblock Accessed: 2026-01-15.

\bibitem[{Narayanan~Venkit et~al.(2025)Narayanan~Venkit, Laban, Zhou, Mao, and Wu}]{narayanan2025search}
Narayanan~Venkit, P.; Laban, P.; Zhou, Y.; Mao, Y.; and Wu, C.-S. 2025.
\newblock Search Engines in the AI Era: A Qualitative Understanding to the False Promise of Factual and Verifiable Source-Cited Responses in LLM-based Search.
\newblock In \emph{Proceedings of the 2025 ACM Conference on Fairness, Accountability, and Transparency}, 1325--1340.

\bibitem[{Pennycook et~al.(2020)Pennycook, Bear, Collins, and Rand}]{pennycook2020implied}
Pennycook, G.; Bear, A.; Collins, E.~T.; and Rand, D.~G. 2020.
\newblock The implied truth effect: Attaching warnings to a subset of fake news headlines increases perceived accuracy of headlines without warnings.
\newblock \emph{Management science}, 66(11): 4944--4957.

\bibitem[{Pf{\"a}nder and Altay(2025)}]{pfander2025spotting}
Pf{\"a}nder, J.; and Altay, S. 2025.
\newblock Spotting false news and doubting true news: a systematic review and meta-analysis of news judgements.
\newblock \emph{Nature human behaviour}, 9(4): 688--699.

\bibitem[{Porter and Wood(2022)}]{porter2022political}
Porter, E.; and Wood, T.~J. 2022.
\newblock Political misinformation and factual corrections on the Facebook news feed: Experimental evidence.
\newblock \emph{The Journal of Politics}, 84(3): 1812--1817.

\bibitem[{Rathi, Jurafsky, and Zhou(2025)}]{rathi2025humans}
Rathi, N.; Jurafsky, D.; and Zhou, K. 2025.
\newblock Humans overrely on overconfident language models, across languages.
\newblock \emph{arXiv preprint arXiv:2507.06306}.

\bibitem[{Renault, Mosleh, and Rand(2025)}]{renault2025grok}
Renault, T.; Mosleh, M.; and Rand, D. 2025.
\newblock @ Grok Is This True? LLM-Powered Fact-Checking on Social Media.

\bibitem[{Robertson, Mour{\~a}o, and Thorson(2020)}]{robertson2020uses}
Robertson, C.~T.; Mour{\~a}o, R.~R.; and Thorson, E. 2020.
\newblock Who uses fact-checking sites? The impact of demographics, political antecedents, and media use on fact-checking site awareness, attitudes, and behavior.
\newblock \emph{The International Journal of Press/Politics}, 25(2): 217--237.

\bibitem[{Salvi, Cuevas, and Ribeiro(2026)}]{salvi2026commercialpersuasionaimediatedconversations}
Salvi, F.; Cuevas, A.; and Ribeiro, M.~H. 2026.
\newblock Commercial Persuasion in AI-Mediated Conversations.
\newblock arXiv:2604.04263.

\bibitem[{Sathish et~al.(2024)Sathish, Lin, Kamath, and Nyayachavadi}]{sathish2024llempower}
Sathish, V.; Lin, H.; Kamath, A.~K.; and Nyayachavadi, A. 2024.
\newblock Llempower: Understanding disparities in the control and access of large language models.
\newblock \emph{arXiv preprint arXiv:2404.09356}.

\bibitem[{Smith-Vaniz et~al.(2025)Smith-Vaniz, Lyon, Steigner, Armstrong, and Mattei}]{smith2025investigating}
Smith-Vaniz, N.; Lyon, H.; Steigner, L.; Armstrong, B.; and Mattei, N. 2025.
\newblock Investigating Political and Demographic Associations in Large Language Models Through Moral Foundations Theory.
\newblock In \emph{Proceedings of the AAAI/ACM Conference on AI, Ethics, and Society}, volume~8, 2419--2430.

\bibitem[{Sundar(2020)}]{sundar2020rise}
Sundar, S.~S. 2020.
\newblock Rise of machine agency: A framework for studying the psychology of human--AI interaction (HAII).
\newblock \emph{Journal of computer-mediated communication}, 25(1): 74--88.

\bibitem[{Weeks(2015)}]{weeks2015emotions}
Weeks, B.~E. 2015.
\newblock Emotions, partisanship, and misperceptions: How anger and anxiety moderate the effect of partisan bias on susceptibility to political misinformation.
\newblock \emph{Journal of communication}, 65(4): 699--719.

\bibitem[{Wood and Porter(2019)}]{wood2019elusive}
Wood, T.; and Porter, E. 2019.
\newblock The elusive backfire effect: Mass attitudes’ steadfast factual adherence.
\newblock \emph{Political Behavior}, 41(1): 135--163.

\bibitem[{Xiong et~al.(2023)Xiong, Lee, Seo, and Lee}]{xiong2023effects}
Xiong, A.; Lee, S.; Seo, H.; and Lee, D. 2023.
\newblock Effects of associative inference on individuals’ susceptibility to misinformation.
\newblock \emph{Journal of Experimental Psychology: Applied}, 29(1): 1.

\bibitem[{Yang and Menczer(2025)}]{yang2025accuracy}
Yang, K.-C.; and Menczer, F. 2025.
\newblock Accuracy and political bias of news source credibility ratings by large language models.
\newblock In \emph{Proceedings of the 17th ACM Web Science Conference 2025}, 127--137.

\bibitem[{Yaqub et~al.(2020)Yaqub, Kakhidze, Brockman, Memon, and Patil}]{yaqub2020effects}
Yaqub, W.; Kakhidze, O.; Brockman, M.~L.; Memon, N.; and Patil, S. 2020.
\newblock Effects of credibility indicators on social media news sharing intent.
\newblock In \emph{Proceedings of the 2020 chi conference on human factors in computing systems}, 1--14.

\bibitem[{Zeff(2023)}]{Zeff2023Grok}
Zeff, M. 2023.
\newblock Enter 'Grok,' Elon Musk's Anti-Woke Chatbot.
\newblock Archived from the original on December 2, 2023; Retrieved December 2, 2023.

\bibitem[{Zhou et~al.(2024)Zhou, Sharma, Zhang, and Althoff}]{zhou2024correcting}
Zhou, X.; Sharma, A.; Zhang, A.~X.; and Althoff, T. 2024.
\newblock Correcting misinformation on social media with a large language model.
\newblock \emph{arXiv preprint arXiv:2403.11169}.

\bibitem[{Zhou and Zafarani(2020)}]{zhou2020survey}
Zhou, X.; and Zafarani, R. 2020.
\newblock A survey of fake news: Fundamental theories, detection methods, and opportunities.
\newblock \emph{ACM Computing Surveys (CSUR)}, 53(5): 1--40.

\end{thebibliography}

\section{Paper Checklist}

\begin{enumerate}

\item For most authors...
\begin{enumerate}
    \item  Would answering this research question advance science without violating social contracts, such as violating privacy norms, perpetuating unfair profiling, exacerbating the socio-economic divide, or implying disrespect to societies or cultures?
    \answerYes{Yes, the research questions answered in this paper advance research on LLMs in the context of fact-checking, without violating social contracts. All data collected through Prolific was done so anonymously with IRB approval and with fair pay. Further, given the use of real, fact-checked information, we ensure that all participants are debriefed about each headlines’s veracity after the experiment.}
  \item Do your main claims in the abstract and introduction accurately reflect the paper's contributions and scope?
    \answerYes{Yes, the claims made throughout the paper have been carefully contextualized. We have made sure that the claims made in the abstract and the introduction accurately reflect the results of the paper.}
   \item Do you clarify how the proposed methodological approach is appropriate for the claims made? 
    \answerYes{Yes, we discuss our methods and justification at length in the Methods section of the paper.}
   \item Do you clarify what are possible artifacts in the data used, given population-specific distributions?
    \answerYes{Yes, we worked to make our data sample and experimental setup clear. We have also provided all materials through OSF.}
  \item Did you describe the limitations of your work?
    \answerYes{Yes, we describe the limitations of our work, as well as ways in which future work can address these limitations.}
  \item Did you discuss any potential negative societal impacts of your work?
    \answerYes{While we did not have space in the paper to thoroughly discuss the potential negative societal impacts of LLM fact-checking, we do discuss it briefly in the discussion.}
      \item Did you discuss any potential misuse of your work?
    \answerNo{No. In this paper, we do not curate or release any resources or software that have potential for misuse. We do not create any artifacts that can be used outside of this work.}
    \item Did you describe steps taken to prevent or mitigate potential negative outcomes of the research, such as data and model documentation, data anonymization, responsible release, access control, and the reproducibility of findings?
    \answerYes{Yes. Participants provided consent to take part in the research. All data was anonymized. We have worked to provide sufficient details for our work to replicated.}
  \item Have you read the ethics review guidelines and ensured that your paper conforms to them?
    \answerYes{Yes, we have reviewed the ethics review guidelines and ensured that this paper conforms to them.}
\end{enumerate}

\item Additionally, if your study involves hypotheses testing...
\begin{enumerate}
  \item Did you clearly state the assumptions underlying all theoretical results?
    \answerNA{NA}
  \item Have you provided justifications for all theoretical results?
    \answerNA{NA}
  \item Did you discuss competing hypotheses or theories that might challenge or complement your theoretical results?
    \answerNA{NA}
  \item Have you considered alternative mechanisms or explanations that might account for the same outcomes observed in your study?
    \answerNA{NA}
  \item Did you address potential biases or limitations in your theoretical framework?
    \answerNA{NA}
  \item Have you related your theoretical results to the existing literature in social science?
    \answerNA{NA}
  \item Did you discuss the implications of your theoretical results for policy, practice, or further research in the social science domain?
    \answerNA{NA}
\end{enumerate}

\item Additionally, if you are including theoretical proofs...
\begin{enumerate}
  \item Did you state the full set of assumptions of all theoretical results?
    \answerNA{NA}
	\item Did you include complete proofs of all theoretical results?
    \answerNA{NA}
\end{enumerate}

\item Additionally, if you ran machine learning experiments...
\begin{enumerate}
  \item Did you include the code, data, and instructions needed to reproduce the main experimental results (either in the supplemental material or as a URL)?
    \answerNA{NA}
  \item Did you specify all the training details (e.g., data splits, hyperparameters, how they were chosen)?
    \answerNA{NA}
     \item Did you report error bars (e.g., with respect to the random seed after running experiments multiple times)?
    \answerNA{NA}
	\item Did you include the total amount of compute and the type of resources used (e.g., type of GPUs, internal cluster, or cloud provider)?
    \answerNA{NA}
     \item Do you justify how the proposed evaluation is sufficient and appropriate to the claims made? 
    \answerYes{Yes. While we did not run machine learning experiments in this paper, we did justified the choice of statistical model based on the structure of the data collected.}
     \item Do you discuss what is ``the cost`` of misclassification and fault (in)tolerance?
    \answerNA{NA}
  
\end{enumerate}

\item Additionally, if you are using existing assets (e.g., code, data, models) or curating/releasing new assets, \textbf{without compromising anonymity}...
\begin{enumerate}
  \item If your work uses existing assets, did you cite the creators?
    \answerNA{NA}
  \item Did you mention the license of the assets?
    \answerNA{NA}
  \item Did you include any new assets in the supplemental material or as a URL?
    \answerNA{NA}
  \item Did you discuss whether and how consent was obtained from people whose data you're using/curating?
    \answerNA{NA}
  \item Did you discuss whether the data you are using/curating contains personally identifiable information or offensive content?
    \answerNA{NA}
\item If you are curating or releasing new datasets, did you discuss how you intend to make your datasets FAIR?
\answerNA{NA}
\item If you are curating or releasing new datasets, did you create a Datasheet for the Dataset? 
\answerNA{NA}
\end{enumerate}

\item Additionally, if you used crowdsourcing or conducted research with human subjects, \textbf{without compromising anonymity}...
\begin{enumerate}
  \item Did you include the full text of instructions given to participants and screenshots?
    \answerYes{Yes. We describe the full experimental flow in the paper. We also provide all stimuli and screenshots throughout the experiment in the supplemental materials.}
  \item Did you describe any potential participant risks, with mentions of Institutional Review Board (IRB) approvals?
    \answerYes{Yes. Our study was IRB approved and described potential participant risks before informed consent was given. The study had minimal risks to participants and all participants were required to complete a debriefing after the experiment.}
  \item Did you include the estimated hourly wage paid to participants and the total amount spent on participant compensation?
    \answerYes{Yes. We include information about participant compensation in the methods.}
   \item Did you discuss how data is stored, shared, and deidentified?
   \answerYes{Yes. Data was all anonymously collected and stored.}
\end{enumerate}

\end{enumerate}

\end{document}